\documentclass[twocolumn,showpacs,prb]{revtex4}
\usepackage{amsfonts}
\usepackage{amsmath}
\usepackage{amssymb}
\usepackage{graphicx}
\usepackage{natbib}
\usepackage{color}
\usepackage[colorlinks=true, letterpaper=true, pdfstartview=FitV, linkcolor=blue, citecolor=blue, urlcolor=blue]{hyperref}

\begin{document}

\title{Tuning electrical and optical anisotropy of a monolayer black
phosphorus magnetic superlattice }
\author{X. J. Li,$^1$ J. H. Yu,$^2$ Z. H. Wu,$^2\dag$ W. Yang$^3\ddag$ \footnote{$^{\dag}$wuzhenhua@ime.ac.cn\\$^{\ddag}$ wenyang@csrc.ac.cn.}}
\affiliation{$^{1}$College of Physics and Energy, Fujian Normal University, Fuzhou
350117, China}
\affiliation{$^{2}$Key Laboratory of Microelectronic Devices and Integrated Technology,
Institute of Microelectronics, Chinese Academy of Sciences, Beijing 100029,
P. R. China}
\affiliation{$^{3}$Beijing Computational Science Research Center, Beijing 100094, P. R.
China}

\begin{abstract}
We investigate theoretically the effects of modulated periodic perpendicular
magnetic fields on the electronic
states and optical absorption spectrum in a monolayer black phosphorus
(phosphorene). We demonstrate that different phosphorene magnetic
superlattice (PMS) orientations can give rise to distinct energy spectra,
i.e., tuning the intrinsic electronic anisotropy. The Rashba spin-orbit coupling (RSOC) will develop a
spin-splitting energy dispersion in this phosphorene magnetic supperlattice.
Anisotropic momentum-dependent carrier distributions along/perpendicular to
the magnetic strips are demonstrated, and the manipulations of these exotic
properties by tuning superlattice geometry, magnetic field and the RSOC term
(via an external electric field) are addressed systematically. Accordingly,
we find bright-to-dark transitions in the ground state electron-hole pairs
transition rate spectrum and PMS orientation dependent anisotropic optical
absorption spectrum. This feature offers us a practical way to modulate the
electronic anisotropy in phosphorene by magnetic superlattice configurations
and detect these modulation capability by using the optical technique.
\end{abstract}

\maketitle

\section{Introduction}

Two dimensional (2D) semiconductor materials have unique layer structure, in
which each layer is vertically stacked by van der Waals force. As a result,
the crystal can be scaled down to atomic layer scale with significant
changes in the physical properties. The changeable bandgap and high mobility
in some 2D materials offer exciting opportunities for development of high
performance electronic and optical devices. To develop these fascinating
applications based on their unique electronic and optical properties, a
scheme of superlattice to manipulate the electronic states and charge flows
in a 2D material based nanostructure has attracted an increasing research
interest. Since the work by Esaki and Tsu,~\cite{Esaki} a great deal of
attention has been devoted to superlattice graphene, where external
spatially periodic electric~\cite{C,Park,Brey,Barbier,Maksimova,ZWu,Uddin}
and/or magnetic fields are applied to a graphene monolayer.~\cite%
{Anna0,Le,Anna,Uddin2,XLi} In these previous studies, people have
demonstrated effective band engineering and optical modulation by the real
superlattice structure as well as periodic external fields.

\begin{figure}[tbp]
\includegraphics[width=\columnwidth]{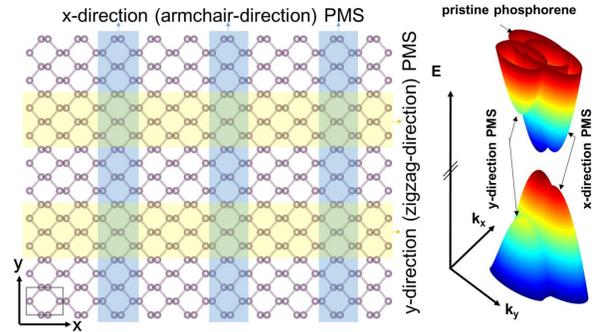}
\caption{(Left) Schematic diagram of a PMS with periodic magnetic stripes along the $x$ direction (armchair
direction) and the $y$ direction (zigzag direction) respectively.
(Right) Strongly anisotropic energy spectra of pristine phosphorene, $x$-direction PMS, and $y$-direction PMS.}
\label{fig1}
\end{figure}

Recently, black phosphorus(BPs), a rare allotrope of phosphorus, is one
of a new type of 2D materials. The high electronic mobility has already been
confirmed in few-layer BPs and applied to the applications of field-effect
transistors (FET) \cite{Li}. The layer structure of BPs held together by
van-der-Waals forces can be exfoliated from few layers to monolayer
(phosphorene), with the layer-dependent direct band gap from $0.3$ eV \cite%
{Morita} (bulk) to $1.52$ eV \cite{Liu} (monolayer), leading to potential
applications in optoelectronics, especially in the infrared regime. 
Interestingly the concept of utilizing strong anisotropic properties of 2D
materials for novel optoelectronic and electronic device applications has
been proposed. Strongly anisotropic conducting behavior~\cite{Jingsi},
anisotropic exciton~\cite{XWang,VyTran} or optical response~\cite{Xia,YXie},
anisotropic Landau levels~\cite{XZhou4}, anisotropic Rashba spin-orbit coupling~\cite{Popovic} as well as anisotropic structural flexibility~\cite{GWang}, have been observed. Inspired by the
supperlattice structure where external spatially periodic electric and/or
magnetic fields are applied to a graphene monolayer, people pay attention to
the supperlattice BPs with varying external periodic potentials and the dependence of BP anisotropy on the periodic perturbation. The
transmission probability of the wave packets with normal incidence can be
tuned and controlled.\cite{zhengli} An analytical model is presented to
relate the decrease in the direct band gap to the different orbital
characters between the valence conduction band.~\cite{Ono} Beside electric controlling,
the effects of magnetic fields on BP's electronic and optical properties have also been studied.~\cite{RZhang,XZhou2,XZhou3,LLi,RZhang2}
However periodic magnetic
modulations, especially for the experimental measurable quantities,
like conductance, optical absorption spectrum, have not been investigated
thoroughly until now.

In this work, we propose a monolayer black phosphorus (or phosphorene) based
PMS, which is an ideal system to archive anisotropic two-dimensional
electron system with high flexibility and controllability. We show that the
anisotropic energy dispersions of the phosphorene can be effectively tuned
by different magnetic stripes configurations, the strength of the magnetic
fields as well as the RSOC interactions. Accordingly we illustrate the
impacts of such a periodic magnetic field on the anisotropic
transport properties including the effective mass. We moreover demonstrate
that the corresponding energy states of e-h pairs and optical transition
rates can also be controlled by the periodic magnetic stripes. Besides the
crystalline induced anisotropy along the two real-space orthogonal axes, the
magnetic fields break the time-reversal symmetry and develop an in-line
anisotropy along the reciprocal-space axes, \emph{e.g.}, the bright-to-dark
transition in optical transition rate of e-h pairs which also can be tuned
by different magnetic stripe configurations and the RSOC interactions.

\section{Theoretical model}

We consider phosphorene(BP monolayer) coated by periodic magnetic stripes
along the plane direction as shown in Fig.~\ref{fig1}. The magnetic field is applied
perpendicular to the layer. The low-energy dispersion of bulk BP can be well described
by a two band $\vec{k}\cdot\vec{p}$ effective mass Hamiltonian due to the $D_{2h}$
point group invariance~\cite{LVoon},
\begin{equation}
H=H_{0}(\mathbf{p}+e\mathbf{A})+H_{R}+H_{Z},
\end{equation}
where $H_{0}$, $H_{R}$, $H_{Z}$ are the two-band effective Hamiltonian, the RSOC term, the
Zeeman term respectively. The first term is given by,~\cite{RFei,YJiang,XZhou}
\begin{equation}
H_{0}=
\begin{bmatrix}
E_{c}+\alpha_{c} k^2_{x}+ \beta_{c} k^2_{y} & \gamma k_{x} \\
\gamma k_{x} & E_{v}+\alpha_{v} k^2_{x}+ \beta_{v} k^2_{y}%
\end{bmatrix}.
\label{H}
\end{equation}
The band parameters are $\alpha _{c}=\hbar ^{2}/2m_{cx}$, $\beta _{c}=\hbar
^{2}/2m_{cy}$, $\alpha _{v}=\hbar ^{2}/2m_{vx}$, $\beta _{v}=\hbar
^{2}/2m_{vy}$, $m_{cx}=0.793m_{e}$, $m_{cy}=0.848m_{e}$, $m_{vx}=1.363m_{e}$%
, $m_{vy}=1.142m_{e}$. The $x$-direction periodic PMS induce periodic
magnetic fields of pointing up and down sequentially. Accordingly the vector
potential $\mathbf{A}$ in each cell are given by,
\begin{eqnarray}
\mathbf{B} &=&\left\{
\begin{array}{c}
(0,0,B);\text{ \ \ \ \ \ \ \ \ \ \ }x\in \lbrack -\frac{W}{2},0) \\
(0,0,-B);\text{ \ \ \ \ \ \ \ \ \ \ \ }x\in \lbrack 0,\frac{W}{2}] \\
(0,0,0);\text{ \ \ \ \ \ \ \ \ \ }x\notin \lbrack -\frac{W}{2},\frac{W}{2}]%
\end{array}%
\right.    \\
\mathbf{A} &=&\left\{
\begin{array}{c}
(0,Bx+\frac{WB}{2},0);\text{ \ \ \ }x\in \lbrack -\frac{W}{2},0) \\
(0,-Bx+\frac{WB}{2},0);\text{ \ \ \ }x\in \lbrack 0,\frac{W}{2}] \\
(0,0,0);\text{ \ \ \ \ \ \ \ \ \ }x\notin \lbrack -\frac{W}{2},\frac{W}{2}]%
\end{array}%
\right. \label{A_Xlatt}
\end{eqnarray}%
in which $W$ is the width of magnetic stripes. Similar form for $y$%
-direction periodic PMS,
\begin{eqnarray}
\mathbf{B} &=&\left\{
\begin{array}{c}
(0,0,-B);\text{ \ \ \ \ \ \ \ \ \ \ }y\in \lbrack -\frac{W}{2},0) \\
(0,0,B);\text{ \ \ \ \ \ \ \ \ \ \ \ }y\in \lbrack 0,\frac{W}{2}] \\
(0,0,0);\text{ \ \ \ \ \ \ \ \ \ }y\notin \lbrack -\frac{W}{2},\frac{W}{2}]%
\end{array}%
\right.    \\
\mathbf{A} &=&\left\{
\begin{array}{c}
(By+\frac{WB}{2},0,0);\text{ \ \ \ }x\in \lbrack -\frac{W}{2},0) \\
(-By+\frac{WB}{2},0,0);\text{ \ \ \ }x\in \lbrack 0,\frac{W}{2}] \\
(0,0,0);\text{ \ \ \ \ \ \ \ \ \ }x\notin \lbrack -\frac{W}{2},\frac{W}{2}]%
\end{array}%
\right. \label{A_Ylatt}
\end{eqnarray}
Then we include only the most important anisotropic Rashba SOC term which is
linear in momentum as given by~\cite{Popovic},
\begin{widetext}
\begin{equation}
H_{R}=
\begin{bmatrix}
0 & 0 & -i(R_{cx}k_{x}-iR_{cy}k_{y}) & 0 \\
0 & 0 & 0& -i(R_{vx}k_{x}-iR_{vy}k_{y}) \\
i(R_{cx}k_{x}+iR_{cy}k_{y}) & 0 & 0 &0 \\
0 & i(R_{vx}k_{x}+iR_{vy}k_{y}) & 0 & 0
\end{bmatrix}
\label{HRashba}
\end{equation}
\end{widetext}
in which $R_{cx}=0.014eV\dot{A}$, $R_{cy}=0.0017eV\dot{A}$, $R_{vx}=0.0109eV%
\dot{A}$, and $R_{vy}=0.0036eV\dot{A}$.~\cite{Popovic} The last term $%
H_{Z}=g\mu _{B}\mathbf{\sigma \cdot B}$ represents the familiar Zeeman
effect. And therefore our full Hamiltonian is $H=H_{0}+H_{R}+H_{Z}$.

\begin{figure*}[tbp]
\includegraphics[width=1.8\columnwidth]{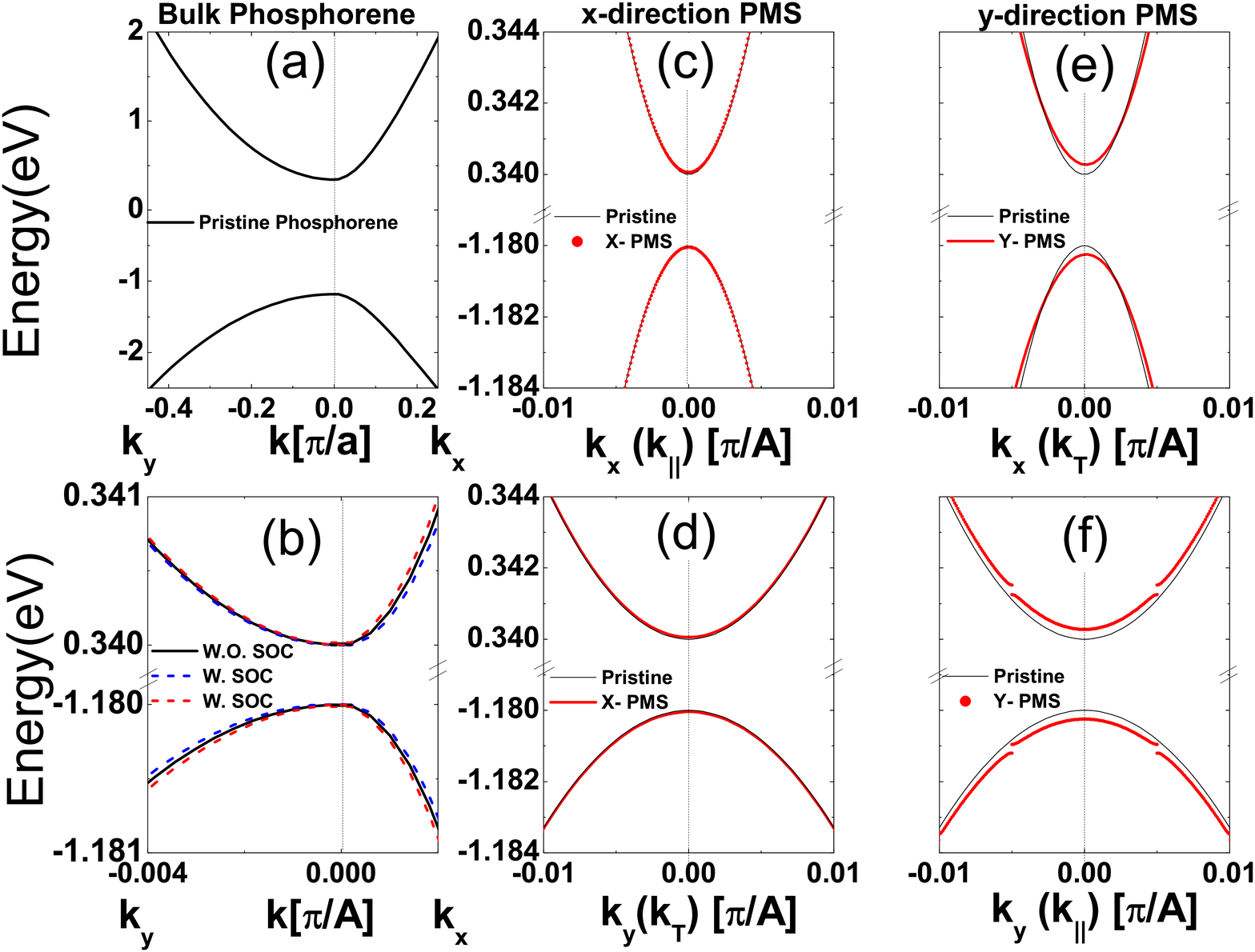}
\caption{(a) The low energy dispersions of pristine phosphorene without RSOC in a wide range; (b) The same as (a), but with RSOC in the vicinity of $\Gamma$ point;
(c) The low energy dispersions of a x-direction PMS along the periodic direction with $k_{\top}$ = 0; (d) The same as (c), but along the transverse direction with $k_{||}$ = 0;
(e) The low energy dispersions of a y-direction PMS,along the transverse direction with $k_{||}$ = 0; (f) The same as (e) but along the periodic direction $k_{\top}$ = 0.  $k_{||}$ ($k_{\top}$) denote the wave vectors along (perpendicular to) the periodic direction, respectively. The PMS parameters are fixed with $B=1$ T, $W=10$ nm, $L=20$ nm. For (d) and (e) the band minima are move back to the $\Gamma$ point for paired comparison analysis.}
\label{fig2_E_k}
\end{figure*}

The envelope functions may be combined into a four component spinor $\Psi
=(\Psi _{c}\uparrow,\Psi _{v}\uparrow,\Psi _{c}\downarrow,\Psi
_{v}\downarrow)$ which satisfies a Dirac equation $H\Psi =E\Psi$. The
electron wave functions are expanded in a plane wave basis confined by the
large hard wall box. The wave function $\Psi $ for electron can be expanded
as
\begin{equation}
\begin{split}
\Psi(k_{x},k_{y})=\sum_{n}\mathbf{C}_{n}\phi_{n}(k_{x},k_{y}) \\
=\sum_{n}\mathbf{C}_{n}\frac{1}{\sqrt{L}}e^{i(\frac{2n\pi x}{L}%
+k_{x}x)}e^{ik_{y}y},  \label{psi}
\end{split}%
\end{equation}%
where $k_{x}$ ($k_{y}$) is the wave vector in the x (y) direction, and the
expansion coefficient $\mathbf{C}_{n}$ a four-component column vector. The
wave function can be calculated numerically in the basis set with the
periodic boundary conditions in the $x$ ($y$) direction.

The interaction Hamiltonian between the Dirac fermion and the photon within
the electrical dipole approximation is $H_{int}=H(\vec{p}+e\vec{\mathcal{A}})-H(\vec{p}%
)$~, where the vector potential $\vec{\mathcal{A}}=(\mathcal{A}_{x}\pm i\mathcal{A}_{y}%
)e^{-i\omega t}$ corresponds to the $\sigma \pm$ circularly polarized light.
$|i>$ denotes the initial states in the lower cones that are hole or valence
like, $|f>$ denotes the final states in the upper cone states that are
electron of conduction like. Now the electron-light interaction induces
transition from $|i>$ to $|f>$, $|i>$ and $|f>$ are written as $\Psi
_{e,h}(k_{x},k_{y})$. The resulting optical transition rate of e-h pair
between valence and conduction band is $|<f|H_{int}|i>|$. The transition
rate is given by,
\begin{equation}
w_{if}=2\pi \delta (E_{f}-E_{i}-\hbar \omega )|<f|H_{int}|i>|^{2}.
\label{TR}
\end{equation}%
in which $<f|H_{int}|i>=\sum_{n,m}\mathbf{C}^{+}_{f,n}\phi_{f,n}^{*}H_{int}%
\mathbf{C}_{i,m}\phi_{i,m}$. Finally we can obtain the optical absorption
rate by the integral of transition rates in $k$ space,
\begin{equation}
\alpha (\hbar \omega )=\iint_{k_{x},k_{y}}\sum_{i,f}w_{if}dk_{x}dk_{y}.
\label{PL}
\end{equation}

\section{Numerical results and discussions}

In pristine phosphorene parabolic energy bands reflect the existence of massive 2DEG with a direct band gap about $1.52$ eV at the $\Gamma$
point, being different from their counterparts of massless Dirac fermions in another well known 2D material
graphene of hexagonal symmetry. The electron-hole symmetry is broken owing to the interband coupling. The conduction band valence band
coupling effects influence the band structure effectively and lead to
distinct electronic and optical properties rather than being observed in
graphene. We start by investigating the energy dispersion relationships of a
pristine phosphorene monolayer without any external magnetic field or with
periodic magnetic field modulations along the $x$ direction (armchair
direction) and the $y$ direction (zigzag direction) respectively.
As sketched in Fig.~\ref{fig1}, we find strongly PMS direction dependent energy bandgap enlargement, cone shift and dispersion modulation, which we will discuss below in details.
First we plot the energy dispersions of pristine phosphorene in Fig.~\ref{fig2_E_k}(a) and (b). Note that the different slopes of the energy dispersions along two
orthogonal $k_{x}$ and $k_{y}$ directions indicate the anisotropic group
velocities and effective masses in phosphorene (see Fig.~\ref{fig2_E_k}(a)). This unique feature is also
absent in graphene. When the RSOC interaction is incorporated into the
calculations, we can find the expected spin splitting in the conduction and
valence bands as shown by the dashed lines in Fig.~\ref{fig2_E_k}(b).
The spin splitting energy is proportional to the wave vector $k$.
Its impact on the band structure of phosphorene is very small near the $\Gamma$
point on the order of magnitude of 10 $\mu$eV.  The spin splittings due to the RSOC are also antitropic for conduction and valence bands.

Next we investigate the energy dispersions of a PMS with periodic
orientations along the two orthogonal crystal axes, with fixed $B=1$ T, $W=10$ nm, $L=20$ nm.
In Fig.~\ref{fig2_E_k}, we can clearly see that the band minima are shifted away from the $\Gamma$ point along the transverse directions of the two type PMSs due to the
time reversal symmetry breaking.
To obtain a direct observation of energy dispersion modulation by PMS, we move the center of shifted energy cones back to $\Gamma$ point, see Fig.~\ref{fig2_E_k}(c) - (f).
When the periodic magnetic field varies along
the $x$ direction, the energy dispersion of PMS with given magnetic fields is almost unchanged as compared to that of pristine phosphorene.
Since the periodic perturbation on the off-diagonal interband coupling term is absence as indicated by Eq.~\ref{H} and~\ref{A_Xlatt}.
For a $y$-direction PMS, i.e., the periodic
magnetic field varies along the $y$ direction, the energy spectra in the wave
vectors $k_{x}$ and $k_{y}$ are shown in Fig.~\ref{fig2_E_k}(e) and (f).
PMS show peculiar behaviors of gap modulation at the $\Gamma
$ point. In conventional phosphorene superlattice with electric potentials,
the energy gap is reduced~\cite{Ono}. Our proposal of magnetic phosphorene
superlattices, however, are different in that the band gap is increased due
to the external magnetic fields. The slopes of the conduction and valence
energy bands decrease with the existence of magnetic field. Along the periodic direction,
a small gap is opened at the boundary of each superlattice Brillouin zone, e.g. $k_{y}(k_{||})=\pm\pi/L+2n\pi/L$,
due to the perturbation potentials coming from the periodic magnetic fields.
In the transverse direction, the energy spectrum $E[k_{x}(k_{T})]$ becomes slightly asymmetric with respect to $k_{y}=0$, which is absence in phosphorene superlattice with electric potentials. It comes from the vector potentials accounting for time-reversal symmetry breaking by the
magnetic fields, and therefore becomes more distinct as we increase the magnitude or the active area of the magnetic fields as we will discuss later.
This feature is in consistent with previous studies on a single quantum well between two electric/magnetic/strain barriers~\cite{ZWu1,ZWu2,ZWu3}.
For a preliminary summary, the impacts of the PMS with different configurations on the energy
dispersion relationships are quit distinct.  The
energy spectrum in $y$-direction PMS (corresponding to the transport direction along
the zigzag direction) is much more sensitive to the external periodic magnetic
fields than that in its counterpart $x$-direction PMS.

\begin{figure}[tbp]
\includegraphics[width=\columnwidth]{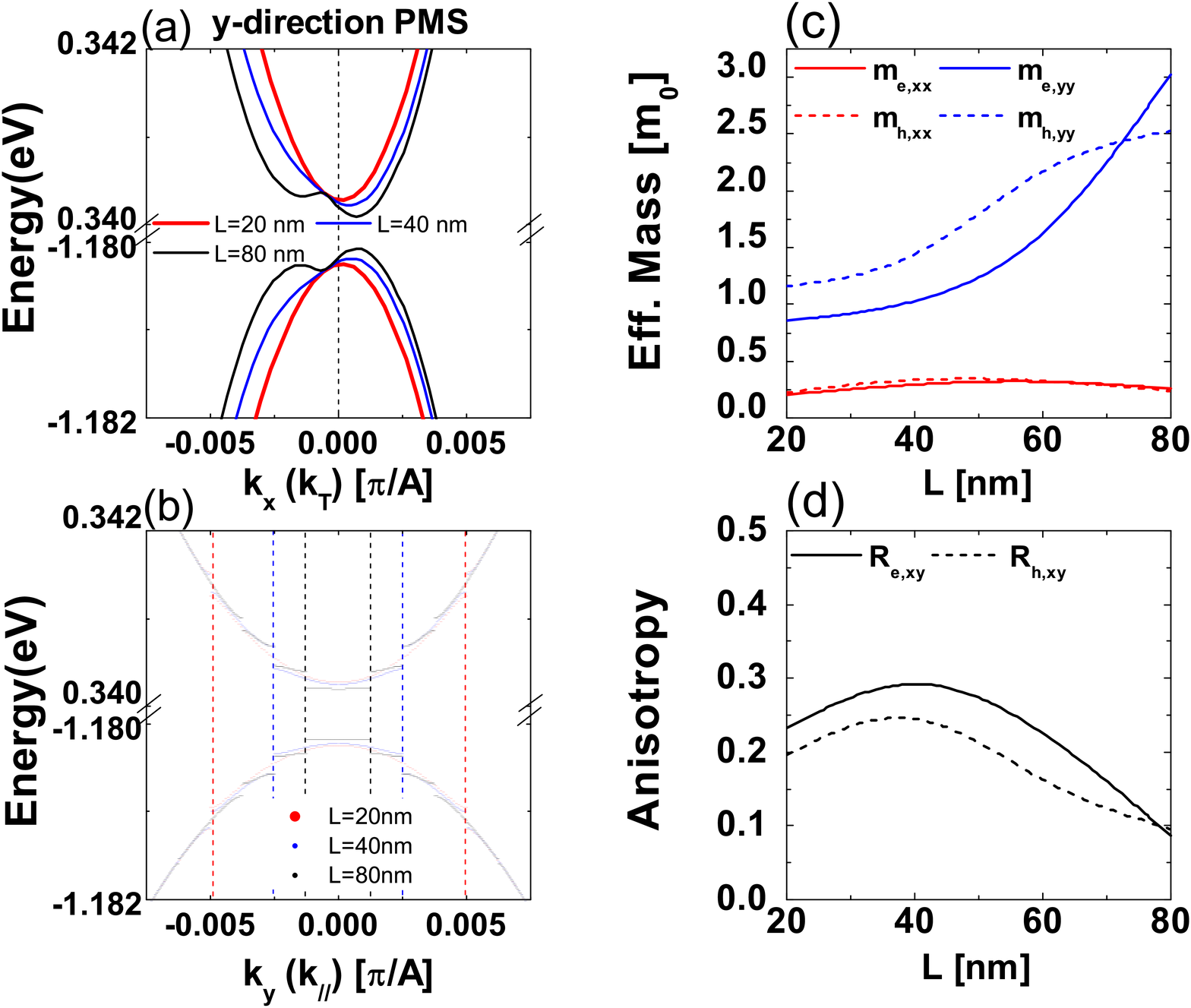}
\caption{(a) and (b) The same as Fig.~\ref{fig2_E_k} (e) and (f), but with different superlattice period lengths.
 (c) and (d) Anisotropic effective mass in $y$-direction PMS as a function of the superlattice period length for both electrons and holes. $R_{e,xy}\equiv m_{e,xx}/m_{e,yy}$, $R_{h,xy}\equiv m_{h,xx}/m_{h,yy}$.}
\label{fig3_E_w}
\end{figure}

\begin{figure}[tbp]
\includegraphics[width=\columnwidth]{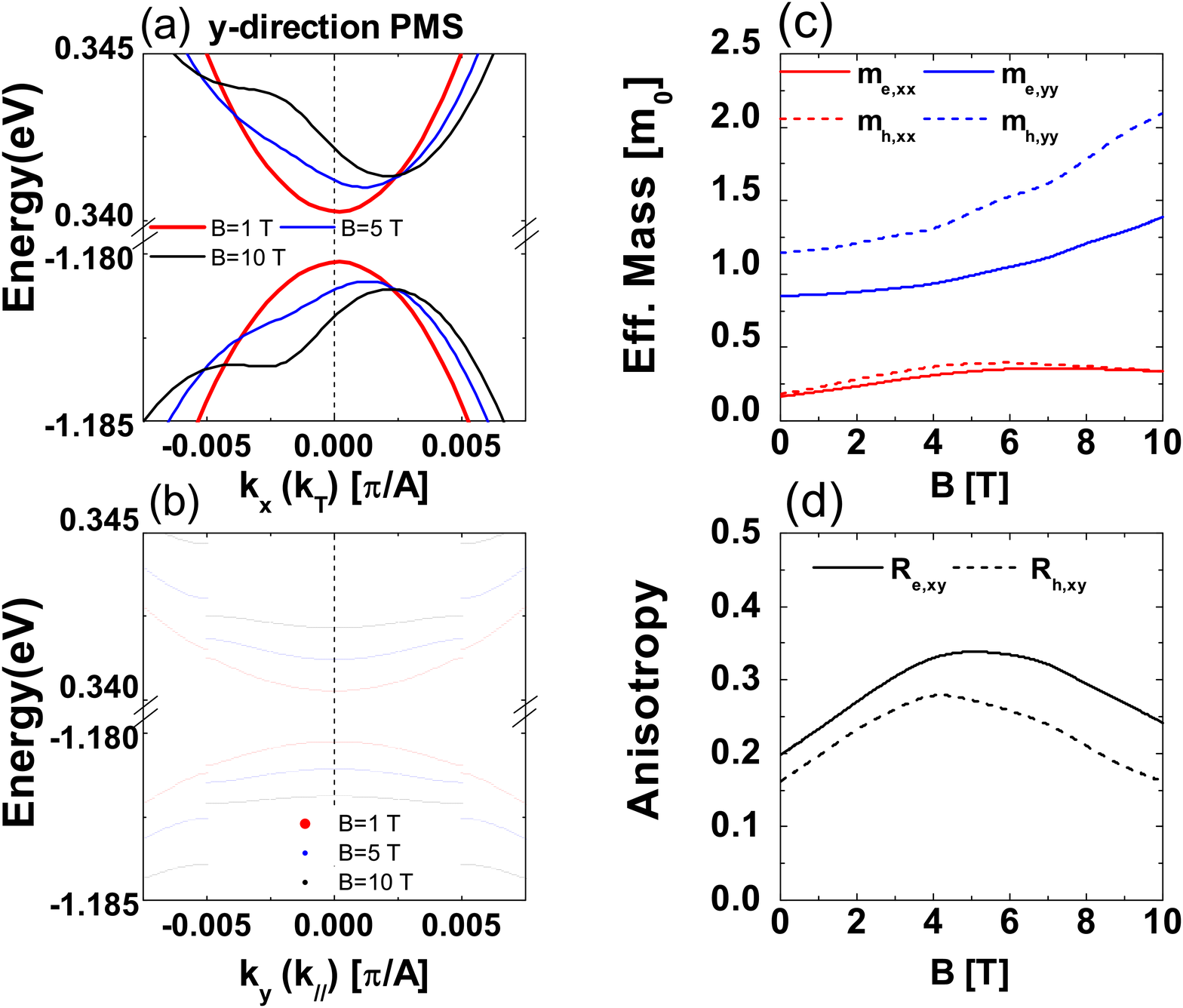}
\caption{(a) and (b) The same as Fig.~\ref{fig2_E_k} (e) and (f), but with different magnetic field strength. (c) and (d) Anisotropic effective mass in $y$-direction PMS as a function of the magnetic field.}
\label{fig4_E_B}
\end{figure}

Then we focus on the energy dispersion and anisotropy modulation in $y$-direction PMS,
since its interband coupling term gives rise to more pronounced response to the periodic vector potential perturbations than that in $x$-direction PMS.
Heuristically, the magnetic fields tend to bend the carriers away from
motion direction and thus suppress the group velocities of charge carriers
as well as reduce energy dispersion. To get an quantitative assessment of
the modulation effects on the electronic properties from the superlattice
structure and magnetic field strength, we then plot the energy dispersions
of $y$-direction PMS along $k_{x}$ and $k_{y}$ with different periodic lengths and
magnetic fields in Figs.~\ref{fig3_E_w} and~\ref{fig4_E_B}. As the
magnetic strip width increases, the aforementioned  bi-directional anisotropy along the transversal
direction (see Fig.~\ref{fig2_E_k}(e)) becomes increasingly spectacular and develops two band minima being adjacent to the $\Gamma$ point as shown in Fig.~\ref{fig3_E_w}(a).
We note en passant that it is a magnetic field induced in-line anisotropy~\cite{ZWu2}, rather than the intrinsic in-plane anisotropy in pristine phosphorene.
Along the periodic direction, more band gaps are opened arising from the periodic perturbation effect at a reduced Brillouin zone boundary.
When the periodic lengths $L$ is very large, the first subband is rather flat along the periodic direction, indicating the presence of heavy quasi particles in such a $y$-direction PMS.
While the energy dispersion along the transversal direction almost remains. Therefore the effective mass and anisotropy can be effectively tuned by the proposed PMS as shown in Fig.~\ref{fig3_E_w}(c) and (d).
Alternatively we can observe similar effects by increasing the strength of the magnetic fields. In addition
band gap enlargement is more pronounced at the $\Gamma$ point and at superlattice Brillouin zone boundaries.
This trend is in contrast with the periodic electrical potentials modulated phosphorene~\cite{zhengli,Ono}.

So, the proposed PMS can effectively tune the anisotropy of energy dispersion and effective mass, since the energy dispersion along the
transversal direction of the superlattice is almost immune to magnetic fields
while the one along the longitude direction of the superlattice is
effectively modulated.
Beside the unique anisotropy between the highly
anisotropic band structures along the two crystalline directions of
phosphorence , in-line bidirectional anisotropy is induced by the magnetic fields.
Importantly, here we address that modulation capabilities of different PMS periodic orientations
are also highly anisotropic as we have discussed about Fig.~\ref%
{fig2_E_k}. For the configuration of $y$-direction (zigzag direction)
PMS, bandstructures are more sensitively depend on the magnetic
fields strength $B$ and periodic cell size $L$.

\begin{figure}[tbp]
\includegraphics[width=\columnwidth]{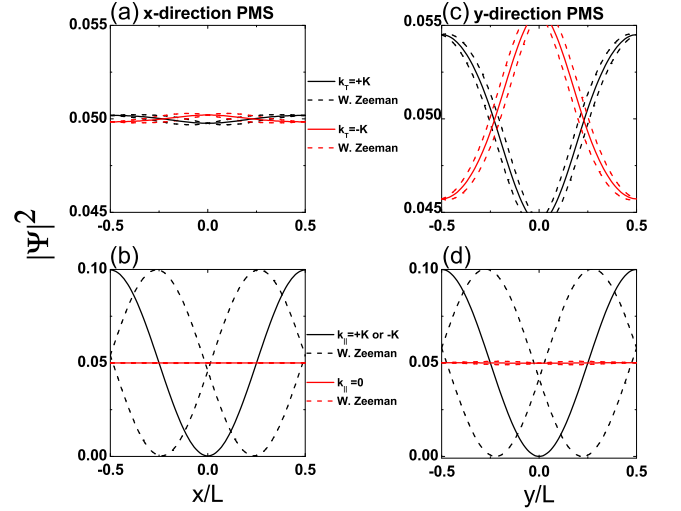}
\caption{The probability distributions of ground-state electrons with specific wave vectors in a superlattice cell, i.e.,
wave vector sampling (a) along the periodic direction, (b) along transverse direction of a $x$-direction PMS,
(c) along periodic direction, and (b) and along transverse direction of a $y$-direction PMS,
$K\equiv\pi/L$ is half the length of the PMS Brillouin zone.}
\label{fig5_distribution_zeeman}
\end{figure}

\begin{figure}[tbp]
\includegraphics[width=\columnwidth]{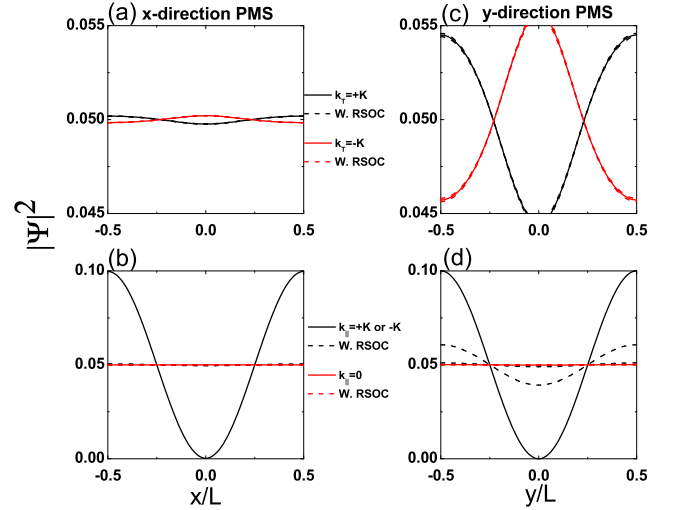}
\caption{The same as Fig.~\ref{fig5_distribution_zeeman}, expect for turning on/off the RSOC term rather than the Zeeman term}
\label{fig6_distribution_RSOC}
\end{figure}

Next we turn our focus to the carrier probability distributions in a unit cell of the PMS
with different configurations. The projected charge distribution of the electrons
in the first conduction band with certain $k_{x}$ and $k_{y}$ are shown in
Fig.~\ref{fig5_distribution_zeeman} and Fig.~\ref{fig6_distribution_RSOC}.
With the wave vector along the magnetic strip (perpendicular to the periodic direction of the
superlattice), the probability distributions exhibit anisotropic dependence
on the wave vector as shown by solid lines in Fig.~\ref%
{fig5_distribution_zeeman} (a) and (c) for $x$-direction periodic and $y$%
-direction periodic superlattice respectively. For a $x$-direction (or $y$%
-direction) PMS, the electrons from the first conduction band with positive $%
k_{y}$ (or $k_{x}$) tend to locate near the edges of a cell where no
magnetic stripe is coated above. While for negative $k_{y}$ (or $k_{x}$), the
electrons tend to locate in the center. This behavior directly arises from
the Hall effect in the proposed PMS. Note that the spatial separation of
electrons moving in the opposite direction is more pronounced with periodic
magnetic modulations along the $y$ direction as shown in Fig.~\ref%
{fig5_distribution_zeeman} (c). This feature is unique in phosphorene, since
the effective vector potential ($A_{x}$) has distinct effects on the off---%
diagonal elements in Hamiltonian Eq.~\ref{H}, while this impact is absence
in $x$-direction PMS. Along the superlattice
direction, the probability distribution are isotropic for positive or
negative wavevectors regardless $x$-direction or $y$-direction
superlattice configurations (see Fig.~\ref{fig5_distribution_zeeman} (b) and
(d)). The electrons tend to distribute in the central region of a cell
underneath the magnetic stripe. Since the perpendicular magnetic fields
develop circular electron orbits and prevent them transmitting away.
By incorporation the Zeeman term, we find that it has negligible impact on
the energy spectrum but change the probability distribution apparently as
shown by dashed lines in Fig.~\ref{fig5_distribution_zeeman}. In the plane
wave basis, the diagonal matrix elements of Zeeman term are canceled out and
the off-diagonal elements are imaginary numbers with the given periodic
magnetic field profile. Therefore the Zeem term can hardly affect the
eigenvalues but changes the eigenvector effectively. The RSOC term can also hardly affects the density distributions with wave vector along the direction of magnetic stripes as shown in Fig.~\ref{fig6_distribution_RSOC} (a) and (c). But it tends to push the electrons with wave vector along the superlattice
direction out of the phosphorene plane resulting in the squeezed
distributions as shown in Fig.~\ref{fig6_distribution_RSOC} (b) and (d).

\begin{figure}[tbp]
\includegraphics[width=\columnwidth]{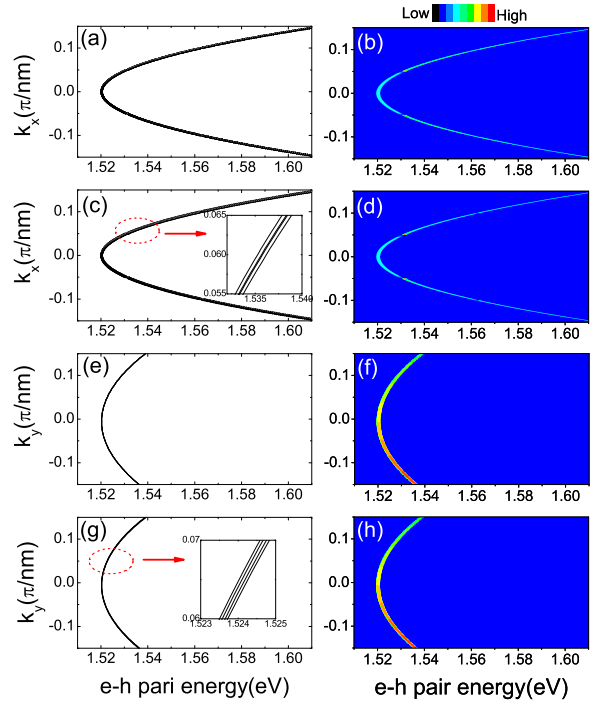}
\caption{(a) The energy dispersions of e-h pairs as a function of wave vector $k_{x}$ with $B=1$ T, $R_{0}=0$
 around the band bottom in a $x$-direction PMS. (b) The optical transition rate of e-h
pairs as a function of wave vector $k_{x}$. (c) and (d) are the same as (a)
and (b) but including the RSOC term. (e)-(h) are the corresponding energy
dispersions and the optical transition rate as a function of the wave vector
$k_{y}$.}
\label{fig7_x_PL}
\end{figure}

\begin{figure}[tbp]
\includegraphics[width=\columnwidth]{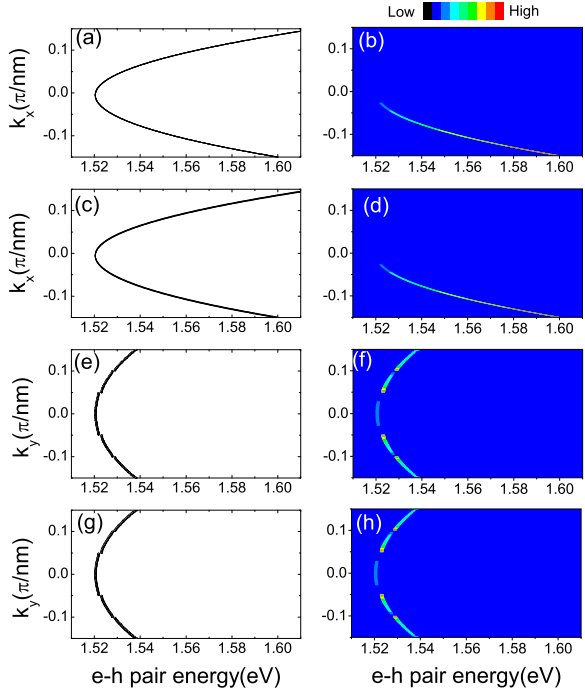}
\caption{The same as Fig.~\ref{fig7_x_PL}, but in a $y$-direction PMS. (a) The energy
dispersions of e-h pairs as a function of wave vector $k_{x}$ with $B=1T$,
$R_{0}=0$ around the band bottom. (b) The optical transition rate of e-h
pairs as a function of wave vector $k_{x}$. (c) and (d) are the same as (a)
and (b) but including the RSOC term. (e)-(h) are the corresponding energy
dispersions and the optical transition rate as a function of the wave vector
$k_{y}$.}
\label{fig8_y_PL}
\end{figure}

The anisotropy energy spectra and charge distributions are hard to measure
directly. However they apparently can affect the Photoluminescence (PL)
Spectra. We therefore investigate the effect of the PMS on the energy to form electron-hole (e-h) pairs by $%
\sigma+$ circularly polarized light and the transition rate. Figs.\ref{fig7_x_PL} and \ref{fig8_y_PL} plot them as functions
of the wave vector $k_{x}$ ($k_{y}$) with periodic magnetic field
modulations along the $x$ ($y$) direction respectively. When considering the
transition between two (accounting for both spin up and spin down states) highest valence bands to two lowest conduction bands,
we can see the parabolic energy spectra of e-h pairs as shown in Fig.~\ref%
{fig7_x_PL} (a), and the transition rates of low energy e-h pairs are
isotropic in $k_{x}$ and rather bright when $k_{x}$ approaches $0$. For
larger $k_{x}$ away from $0$ it becomes dark arising from the electron
distribution in the conduction and valence band with different $k_{x}$ as
shown in Fig.~\ref{fig5_distribution_zeeman} (c). When $k_{x}$ is not zero,
the charge carriers from conduction bands or valence bands tend to locate
along of the edge of a superlattice cell where no magnetic stripe is
deposited above regardless of positive or negative $k_{x}$. It results in
small overlap integral of the wavefunctions and small transition rate in
Fig.~\ref{fig7_x_PL} (b). When $k_{x}$ approaches zero, the charge carriers tend to
distribute equally in a superlattice and thus increase the overlap integral of the
wavefunctions. Then the optical transition rate is enhanced. Next we
examine the effect of the RSOC term, one can see the expected spin-splitting in the e-h
pair energy spectrum (inset in Fig.~\ref{fig7_x_PL} (c)). Accordingly the RSOC term
can affect the electron distribution at the supperlattice BZ boundaries (Fig.~\ref{fig6_distribution_RSOC} (c)) and reduce the transition rate spectrum of the low-energy e-h pairs, so we can find larger transition gap as
shown in Fig.~\ref{fig7_x_PL} (d).
We also plot the e-h pair energy as function of $k_{y}$ in Fig.~\ref%
{fig7_x_PL} (e), the asymmetrical behavior in the e-h pair energy spectrum is
due to the anisotropic energy spectrum as shown in Fig.~\ref{fig2_E_k}. The
optical transition rate spectrum along $k_{y}$ ($k_{T}$) exhibits an in-line asymmetrical behavior with a
dark-to-bright transition when the $k_{y}$ varies from negative to positive.
The electrons with positive $k_{y}$ tend to locate aside the superlattice
cell while the ones with negative $k_{y}$ tend to locate in the center as
shown in Fig. \ref{fig6_distribution_RSOC} (a) and (b). Different
distributions result in different overlap integral of wavefunctions and thus
determine the optical transition rates as we observed. When the RSOC term is
considered, we can find the spin splitting in the e-h pair energy and
optical transition rate spectrum in \ref{fig7_x_PL} (g) and (h). The optical
transition rate is enhanced the same as that in Fig.~\ref{fig7_x_PL} (d) due
to the same reason.

Compared to the $x$-direction periodic PMS, the modulation effects by a $y$-direction PMS are much more pronounced.
The ground state e-h pairs energy as function of $k_{x}$ is asymmetrical
due to the time-reversal symmetry breaking in the presence of a magnetic
field as shown in Fig.~\ref{fig8_y_PL} (a) and spin slitting due to the RSOC
interaction in Fig.~\ref{fig8_y_PL} (c). Because the electron with negative
or positive $k_{x}$ shows separate distributions as shown in Fig.~ \ref%
{fig5_distribution_zeeman} (c) and \ref{fig6_distribution_RSOC} (c), we can
find the dark-to-bright transition of the ground state e-h pair transition
rate spectrum in Fig.~\ref{fig8_y_PL} (b). The RSOC term has small effect on
the transition rate spectrum in $k_{x}$ accounting for its limited impact on
the electron distribution ( see Fig.~\ref{fig6_distribution_RSOC} (b)).
Corresponding e-h pair energy spectrum in $k_{y}$ are symmetrical as shown
in Fig.~\ref{fig8_y_PL} (e). The e-h pairs energy spectrum
and optical transition rate spectrum of $y$-direction periodic superlattice
 possess larger gaps at the supperlattice BZ boundaries as shown Fig.~\ref{fig8_y_PL}, due to the
enhanced interband coupling by the vector potential of the PMS. The optical transition rate spectrum of the
ground e-h pairs is not bright (see Fig.~\ref{fig8_y_PL} (f)) due to the
marginal distribution as shown in \ref{fig6_distribution_RSOC} (d),
regardless whether $k_{y}$ is negative or positive. Although the spin
splitting is shown in energy spectrum but can not been distinguished in the
transition rate spectrum as shown in Figs.~\ref{fig8_y_PL} (g) and (h).
In brief summary, the optical transition rate of $y$-direction PMS is
different from that of $x$-direction PMS as we discussed about
Fig.~\ref{fig7_x_PL}. Obviously the anisotropy can be adjusted by the
strength of the external magnetic field, the periodic length and the RSOC
term. So we can realize an external field controlled magneto-optical device
base on different PMS configurations.

\begin{figure}[tbp]
\includegraphics[width=0.95\columnwidth]{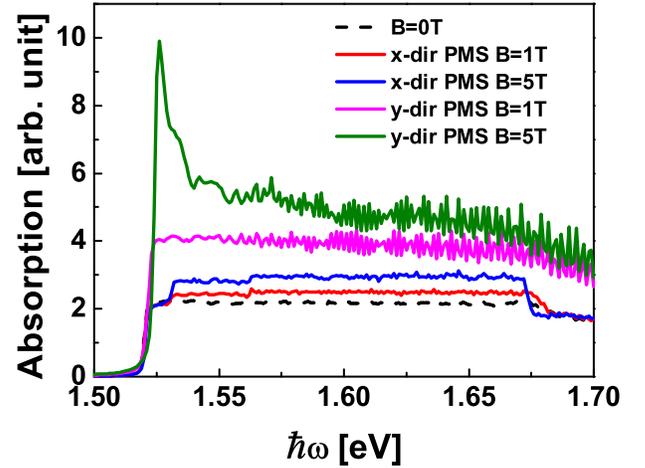}
\caption{The optical absorption spectrum for the phosphorene magnetic
supperlattice with different magnetic stripe configuration and magnetic field. }
\label{fig9_absorption}
\end{figure}

Finally the optical absorption spectrum of such a phosphorene magnetic
supperlattice is calculated which can also be measured directly. In our
calculation, we set the fermi level at zero energy which is between the
conduction and valence band. It means the occupation for the valence band is
full, while that for the conduction band it is empty. We use broadening
factor of $0.15$ $meV$ to smoothen the absorption spectrum. The optical
absorption spectrum indicates useful band structure information guaranteed
by the selection rule expressed as $\delta (E_{f}-E_{i}-\hbar\omega )$ in Eq.%
\ref{TR}. In Fig.~\ref{fig9_absorption}, Different configurations of
periodic magnetic strips show different absorption characteristics. For a $x$%
-direction periodic superlattice, the optical absorption spectrum with one
wide step indicates transition rate enlargement arising from the charge distribution modulations as we discussed before.
For a $y$-direction PMS, we can find much higher absorption peak in low energy region and more oscillations when increasing the energy
(frequency) of the incident light due to modulations from the $y$-direction PMS. Increasing the magnetic field can also increase the absorption rate. The distinct optical absorption
spectra provide an effective way to detect the anisotropic energy properties
of PMS with different periodic orientations. The phosphorene magnetic
supperlattice is a promising platform for potential application in
anisotropic magneto-optical devices.

\section{Conclusions}

In this work, we theoretically investigate the electronic and optical
properties of PMS utilizing the $\vec{k}\cdot\vec{p}$ method. Our numerical
results show that the anisotropic energy dispersions can be tuned by the PMS
configurations, e.g., orientation, periodic length, strength. Accordingly the e-h pair
energies also exhibit distinct differences between two configurations. We demonstrate that $y$-direction
(zigzag direction) superlattice gives rise to more pronounced modulation via
the vector potential appeared in the electron-hole coupling term. As
compared to the energy dispersion, charge distribution is much more
sensitive to the external magnetic field or electric field (via RSOC).
The magnetic fields and periodic orientations of the PMS proposed in this work play important roles in determining the e-h transition rates and optical absorption spectrum. Our
theoretical results shed new light on potential applications of
magneto-optical devices based on the anisotropic PMS.

\begin{acknowledgments}
This work was supported by the MOST (Grants No. 2016YFA0202300) and the
Opening Project of MEDIT, CAS.
\end{acknowledgments}

\end{document}